\documentclass{aa}
\pdfoutput=1
\usepackage{amsmath}
\usepackage[round]{natbib}
\usepackage{graphicx}
\usepackage{flushend}
\usepackage{paralist}
\usepackage[english]{babel}
\usepackage{txfonts}
\usepackage{xcolor}
\usepackage[pdftitle={Limitations for shapelet-based weak-lensing measurements},pdfauthor={Peter Melchior et al.}, bookmarks=true, bookmarksnumbered=true, colorlinks, linkcolor=black,citecolor=black,urlcolor=black]{hyperref}
\bibpunct{(}{)}{;}{a}{}{,} % to follow the A&A style

\newcommand\veqref[1]{Eq. \eqref{#1}}
\newcommand\veqsref[2]{Eqs. \eqref{#1} \& \eqref{#2}}

\newcommand\veqrangeref[2]{Eqs. \eqref{#1} - \eqref{#2}}

\title{Limitations for shapelet-based weak-lensing measurements}
\authorrunning{P. Melchior et al.}
\author{P. Melchior\inst{1}, A. B\"ohnert\inst{2,3}, M. Lombardi\inst{3}, M. Bartelmann\inst{1}}
\institute{Zentrum f\"ur Astronomie Heidelberg, Institut f\"ur Theoretische Astrophysik, Albert-Ueberle-Str. 2, D-69120 Heidelberg, Germany\\
\email{pmelchior@ita.uni-heidelberg.de}
\and Argelander-Institut f\"ur Astronomie, Universit\"at Bonn, Auf dem H\"ugel 71, D-53121 Bonn, Germany
\and European Southern Observatory, Karl-Schwarzschild-Str. 2, D-85748, Garching bei M\"unchen, Germany}
\date{}

\abstract
{% context
}
{% aims
We seek to understand the impact on shape estimators obtained from circular and elliptical shapelet models under two realistic conditions: (a) only a limited number of shapelet modes is available for the model, and (b) the intrinsic galactic shapes are not restricted to shapelet models.
}
{% methods
We create a set of simplistic simulations, in which the galactic shapes follow a S\'{e}rsic profile. By varying the S\'{e}rsic index and applied shear, we quantify the amount of bias on shear estimates which arises from insufficient modeling. Additional complications due to PSF convolution, pixelation and pixel noise are also discussed.
}
{% results
Steep and highly elliptical galaxy shapes cannot be accurately modeled within the circular shapelet basis system and are biased towards shallower and less elongated shapes. This problem can be cured partially by allowing elliptical basis functions, but for steep profiles elliptical shapelet models still depend critically on accurate ellipticity priors. As a result, shear estimates are typically biased low. Independently of the particular form of the estimator, the bias depends on the true intrinsic galaxy morphology, but also on the size and shape of the PSF.
}
{% conclusion
As long as the issues discussed here are not solved, the shapelet method cannot provide weak-lensing measurements with an accuracy demanded by upcoming missions and surveys, unless one can provide an accurate and reliable calibration, specific for the dataset under investigation.
}

\begin{document}
\thispagestyle{empty}
\maketitle

\section{Introduction}
Shapelets have been proposed to describe the complex morphology of galaxies and to compress this information in a small number of expansion coefficients \citep{Refregier03.1}. Indeed, \cite{Massey04.1} find that the shapelet decompositions of deep HST observations preserve the distribution of several morphological estimators so well that one can create simulations of a mock sky which are essentially indistinguishable from actual observations.\\
For weak-lensing measurements it is furthermore important that shapelets can also faithfully describe the PSF shape \citep[e.g.][]{Bernstein02.1,Refregier03.1,Massey05.1,Kuijken06.1}. Additionally, deconvolution from the PSF is an analytic operation in shapelet space \citep{Refregier03.2,Melchior09.1}. This means, given shapelet models of the observed galaxy and the PSF, respectively, only linear algebra is required to obtain a shapelet model of the deconvolved galaxy shape. Finally, several estimators for gravitational shear and flexion can be directly formed from this model \citep{Bernstein02.1,Massey07.2}. In summary, once a shapelet model of each galaxy is obtained, all necessary steps for weak-lensing measurements can be performed very efficiently in shapelet space.

There is, however, an implicit assumption in the setup of such a shapelet-based weak-lensing pipeline: namely that the occurring shapes can be faithfully described by shapelet models. But this assumption is strictly valid only if the basis system is complete, which requires an infinite number of modes. In particular in weak lensing, the amount of modes which can be measured from the images is massively limited. For a limited number of modes, shapes which deviate from the characteristics of the employed basis functions can only have approximate representations. 
 
This limitation applies to any complete basis function set. In this work, we want to understand the consequences particularly for weak-lensing measurements within the shapelet basis. We start by describing the simulated galaxy images in sect. \ref{sec:test_images}. As there are two general shapelet variants, namely circular and -- as a generalization -- elliptical ones, we discuss both of them in sects. \ref{sec:circular_shapelets} and \ref{sec:elliptical_shapelets}. We show the influence of PSF convolution, pixelation and pixel noise in sect. \ref{sec:systematics}, and discuss the consequences of our findings in sect. \ref{sec:conclusions}.

\section{Test images}
\label{sec:test_images}
For this work we seek to answer the following question: How well can a shapelet model recover the shear from a source whose shape is not entirely describable by the model. To pin down the effects of an incomplete shape description and to isolate it from other systematics, we generate very simple test cases in which the galaxy shapes are -- initially -- not affected by PSF convolution and pixel noise.

The radial light profile of galaxies is often described by the S\'{e}rsic profile \citep{Sersic63.1}
\begin{equation}
\label{eq:sersic_profile}
 p_s(r) \propto \exp\Bigl\lbrace -b_{n_s}\Bigl[\Bigl(\frac{r}{R_e}\Bigr)^{1/n_s}-1\Bigr]\Bigr\rbrace,
\end{equation}
where $R_e$ is the radius containing half of the flux\footnote{This is ensured by demanding of $b_{n_s}$ to satisfy the relation  \begin{equation*}\Gamma(2n_s)=2\gamma(2n_s,b_{n_s})\end{equation*} between the complete and the incomplete Gamma function \citep{Graham05.1}. The approximate solution for the equation above, \begin{equation*}b_{n_s}\approx 1.9992 n_s - 0.3271\end{equation*} \citep{Capaccioli89.1} is used throughout the paper.} and $n_s$ is the so-called S\'{e}rsic index. This type of profile is identical to a Gaussian for $n_s=0.5$ and steeper for $n_s > 0.5$. From several investigations it is known that for the vast majority of observable galaxies $n_s$ ranges between 0.5 and 4 \citep[e.g.][]{Sargent07.1}\footnote{The exponential profile which describes the brightness distribution of spiral galaxies and the de Vaucouleurs profile \citep{deVaucouleurs48.1} of elliptical galaxies are special cases of the S\'{e}rsic profile with $n_s=1$ and $n_s=4$, respectively.}.
Thus, we describe the intrinsic shapes $G$ with a flux-normalized S\'{e}rsic profile with $R_e \in \lbrace 5,10,20\rbrace$ pixels and $n_s \in \lbrace 0.5,1,2,3,4\rbrace$. For ensuring finite support, $G$ is truncated at $5\ R_e$.\\

$G$ is sheared in real-space by transforming the coordinates by means of the linearized lens equation \citep[e.g.][]{Bartelmann01.1}
\begin{equation}
 \vec{x} = 
\begin{pmatrix} 
1-\gamma_1 & -\gamma_2 \\
-\gamma_2 & 1+\gamma_1 
\end{pmatrix}
\vec{x^\prime}
\end{equation}
such that $G^\prime(\vec{x^\prime})$ is obtained by evaluating $G$ at the position $\vec{x}$. The values of $\gamma_1$ range between 0 and 0.5; $\gamma_2$ is set to zero.
It is important to notice that $G$ is circular, while observed galaxies show a wide distribution of intrinsic ellipticities \citep{Bernstein02.1}. Hence, $G$ has to acquire its intrinsic ellipticity entirely from the applied shear. To obtain roughly realistic results, the applied shear is varied up to $|\vec\gamma|=0.5$ although such values cannot be generated by the cosmic large-scale structure and are even untypical for all but the innermost parts of galaxy clusters. As an advantage of this procedure, $G$ has elliptical isophotes, for which the axis ratio and orientation are consistent at all radii, and therefore all ellipticity measures formed from these images should agree.

$G^\prime$ is sampled at the final resolution of $20\, R_e\times 20\, R_e$ pixels. While $R_e=5$ is already rather large for typical weak-lensing galaxies, we chose to simulate also even larger ones so as to mimic higher resolution images from which we can assess the impact of pixelation on the shear estimates.

Because there is no pixel noise in these test images and the resolution is very high, the ellipticity $\vec\epsilon$ measured from quadrupole moments of the pixelated image is always compatible with the shear $\vec\gamma$. Additionally, the centroid position can be computed with essentially arbitrary precision from the image.

\section{Circular shapelets}
\label{sec:circular_shapelets}

According to \citet{Refregier03.1},
\begin{equation}
\label{eq:shapelet_basis}
B_\mathbf{n}(\mathbf{x};\beta) \equiv \beta^{-1} \phi_{n_1}(\beta^{-1} x_1)
\ \phi_{n_2}(\beta^{-1}x_2)
\end{equation}
defines the two-dimensional shapelet basis function of order $\vec{n}=(n_1,n_2)$ and scale size $\beta$ which is related to the one-dimensional Gauss-Hermite polynomial
\begin{equation}
 \label{eq:dimless_basis}
\phi_{n}(x) \equiv [2^n \pi^{\frac{1}{2}} n!]^{-\frac{1}{2}}\ H_n(x)\
\mathrm{e}^{-\frac{x^2}{2}},
\end{equation}
with $H_n$ being the Hermite polynomial of order $n$.

From the shapelet coefficients of some two-dimensional function $G$
\begin{equation}
\label{eq:decomposition}
g_\mathbf{n} = \int_{-\infty}^\infty d^2x\ G(\mathbf{x})\ B_\mathbf{n}(\mathbf{x};\beta),
\end{equation}
one can reconstruct a shapelet model
\begin{equation}
\label{eq:model}
\tilde{G}(\mathbf{x}) = \sum_{n_1,n_2=0}^{n_1+n_2=n_{max}} g_\mathbf{n}\, B_\mathbf{n}(\mathbf{x};\beta).
\end{equation}
The model is found by minimizing 
\begin{equation}
\label{eq:chi2}
 \chi^2\propto \sum_{\vec{x}} [G(\vec{x}) - \tilde{G}(\vec{x})]^2
\end{equation}
with respect to the parameters $g_\mathbf{n}$. $\beta$ and $n_{max}$ could be fixed at values chosen to be suitable to describe the galaxy ensemble well on average \citep{Kuijken06.1} or be determined from the minimization of $\chi^2$ as well \citep{Massey05.1,Melchior07.1}.

One foreseeable problem of the shapelet decomposition stems from the Gaussian weighting function in \veqref{eq:dimless_basis}. We discussed already that galaxies typically have steeper profiles than a Gaussian, which means that an optimized shapelet model requires higher orders. However, due to the polynomial in \veqref{eq:dimless_basis}, the largest oscillation amplitudes of high-order modes are located at rather large distances from the centroid. Models which include higher orders thus allow a better description of the outer parts of a galaxy, while they still fail to reproduce correctly the central region in the case of steep profiles. Additionally, in case of noisy image data, the number of modes must be limited to avoid overfitting spurious nearby noise fluctuations. Hence, galactic shapes with steeper profiles than a Gaussian are expected to be described by shapelet models with systematically shallower profiles.

Of similar concern is the circularity of the shapelet basis system. As the scale size for both dimensions in \veqref{eq:shapelet_basis} is the same, the zeroth-order is round. If the shape to be described is stretched in a particular direction -- as a result of its intrinsic shape or due to gravitational lensing -- this elongation has to be carried by higher shapelet orders. Again, for a limited number of basis modes we must expect an insufficient representation of the true shape by the shapelet model. In particular, we have to consider an underestimation of the source elongation much more likely than an overestimation, as the basis system preferentially remains circular.

These shortcomings are likely but do not have to affect the estimation of a weak shear. Whether a model with limited fidelity leads to a biased shear estimate depends an the construction of the estimator.

The most straightforward estimator is obtained from the complex ellipticity \citep[e.g.][]{Bartelmann01.1}
\begin{equation}
 \label{eq:ellipticity}
\vec\epsilon \equiv \frac{Q_{11}-Q_{22}+2iQ_{12}}
   {Q_{11}+Q_{22}+2(Q_{11}Q_{22}-Q_{12}^2)^{\frac{1}{2}}},
\end{equation} 
where $Q_{ij}$ are the quadrupole moments measured from the shapelet model $\tilde{G}^\prime$. From this we obtain a direct shear estimator $\tilde{\vec\gamma}^{(Q)} \equiv \tilde{\vec\epsilon}$.
By choosing this particular definition, we benefit from an ellipticity estimator with a perfect response to external shear, i.e. the shear responsivity $R=\partial\vec\epsilon / \partial\vec\gamma$ \citep[e.g.][]{Bernstein02.1} equals unity and we do not have to correct for it.

It is important to note that the $Q_{ij}$ integrate over the entire shape of $G^\prime$, consequently $\tilde{Q}_{ij}$ are linear combinations of all available shapelet coefficients of $\tilde{G}^\prime$ \citep{shapelets_manual}, therefore this estimator critically relies on a decent shapelet model.

\subsection{Polar shapelets}

For the problem of describing galactic shapes, it is often more convenient to use a polar coordinate frame $(r,\varphi)$ instead of the Cartesian frame $(x_1,x_2)$. Following again \citet{Refregier03.1}, we define the polar shapelet basis functions as
\begin{equation}
\label{eq:polar_shapelets}
B_{n_r,n_l}(r,\varphi)=\beta^{-1}\phi_{n_r, n_l}(\beta^{-1}r,\varphi),
\end{equation}
where $n_l$ and $n_r$ are the left-handed and right-handed modes, respectively, and 
\begin{equation}
\label{eq:polar_basis}
\phi_{n_r, n_l}(r,\varphi) \equiv [\pi n_r! n_l!]^{-\frac{1}{2}}
H_{n_l,n_r}(r) e^{-\frac{x^2}{2}} e^{i(n_r-n_l)\varphi}.
\end{equation}
\citet{Bernstein02.1} showed that, for $n_l<n_r$, one can relate 
\begin{equation}
H_{n_l,n_r}(r) \equiv (-1)^{n_l}n_l! r^{n_r-n_l} L^{n_r-n_l}_{n_l}(r^2)
\end{equation}
to the associated Laguerre polynomial
\begin{equation}
L^q_p(r) \equiv \frac{r^{-q}e^r}{p!}\frac{d^p}{dr^p}(r^{p+q}e^{-r}).
\end{equation}

Note the similarities between the Cartesian basis \veqsref{eq:shapelet_basis}{eq:dimless_basis} and the polar basis \veqsref{eq:polar_shapelets}{eq:polar_basis}: Both share a Gaussian weighting function and are intrinsically circular.

As $B_{n_r,n_l}$ is a complex function if $n_r\neq n_l$, it is computationally more efficient to decompose the galactic shape into Cartesian shapelets according to \veqrangeref{eq:decomposition}{eq:chi2} and then to perform a coordinate transformation $\mathbf{T}^{c\rightarrow p}$ from Cartesian to polar shapelet space \citep[cf. Eq. (69) in][]{Refregier03.1},
\begin{equation}
 g_{n_r,n_l}^p = \mathbf{T}^{c\rightarrow p}\, g_{n_1,n_2},
\end{equation}
We will do that for the tests in the following sections.

Performing this transformation enables us to form a conceptionally different family of shear estimators. By defining $n\equiv n_r + n_l$ and $m\equiv n_r - n_l$, we can see from \veqref{eq:polar_basis}, that $B_{n,m}$ behaves like a field with spin $m$. Since the shear behaves like a spin-2 field, its action on a circular source can be described by the $m=2$ modes of a polar shapelet model. Thus, the most basic one of these estimators,
\begin{equation}
\label{eq:fn2}
 \tilde{\vec\gamma}^{(n2)} \equiv \frac{4}{\sqrt{n(n+2)}}\frac{g^{\prime p}_{n,2}}{\langle g^p_{n-2,0} - g^p_{n+2,0} \rangle},
\end{equation}
uses only the $m=2$ mode of any even polar order $n$ of $\tilde{G}^\prime$ \citep{Massey07.2}\footnote{We restrict ourselves to $n=2$ for the following tests, so that we only discuss $\tilde\gamma^{(22)}$.}. This estimator must be normalized by radial modes $g^p_{n,0}$ obtained from unlensed sources. For the tests discussed here, we provide the unlensed image $G$ such that those coefficients can be measured from images having the correct unlensed shape.

This approach is drastically different from measuring the quadrupole moment -- in the sense that it depends only on a single lensed and two radial unlensed coefficients -- and thus expected to behave in a different manner, even though the decomposition is done with Cartesian shapelets in both cases.

\subsection{Results}
\label{sec:results}

\begin{figure}[t]
 \includegraphics[width=\linewidth]{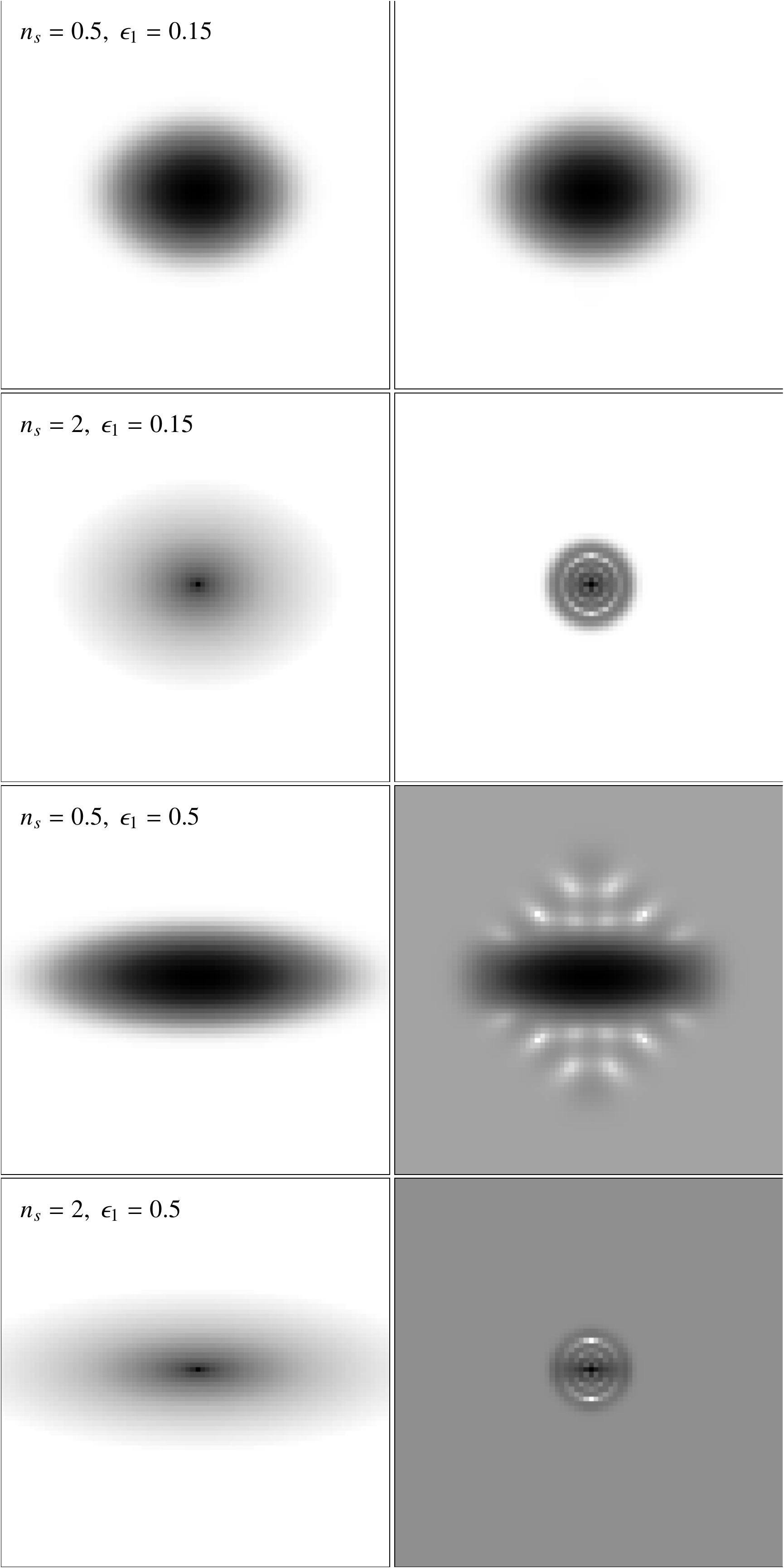}
 \caption{(left) Examples of S\'{e}rsic-type galaxy images with $n_s=0.5$ or 2 and an intrinsic ellipticity (induced by shearing the circular profile \veqref{eq:sersic_profile}) of $\epsilon_1=0.15$ or 0.5. (right) Best-fit circular shapelet models $\tilde{G}^\prime$ with $n_{max}=12$. The color stretch is logarithmic.}
 \label{fig:models}
\end{figure}

$G^\prime$ is decomposed into Cartesian shapelets of maximum order $n_{max}\in\lbrace8,12\rbrace$, which is typical given the significance of weak-lensing images \citep[cf.][]{Kuijken06.1}.

At first, we investigate the modeling fidelity visually. In Fig. \ref{fig:models}, we give four examples of S\'{e}rsic-type galaxy shapes and their shapelet models. It is evident from the top row that for modest ellipticities, an elliptical Gaussian can be very well represented by its shapelet model. But if either the ellipticity becomes stronger or the intrinsic galactic profile becomes steeper, the shapelet decomposition performs poorer. For the Gaussian case shown in the third row, the overall shape is evidently more compact and rather boxy than elliptical, and affected by oscillatory artefacts. The images with $n_s=2$ (second and forth row) show prominent ring-shaped artefacts and are concentrated at the core region of $G$. It is striking that the drastic increase in ellipticity from $\epsilon_1=0.15$ to $\epsilon_1=0.5$ causes almost no apparent change in the respective shapelet models.

\begin{figure}[t]
 \includegraphics[width=\linewidth]{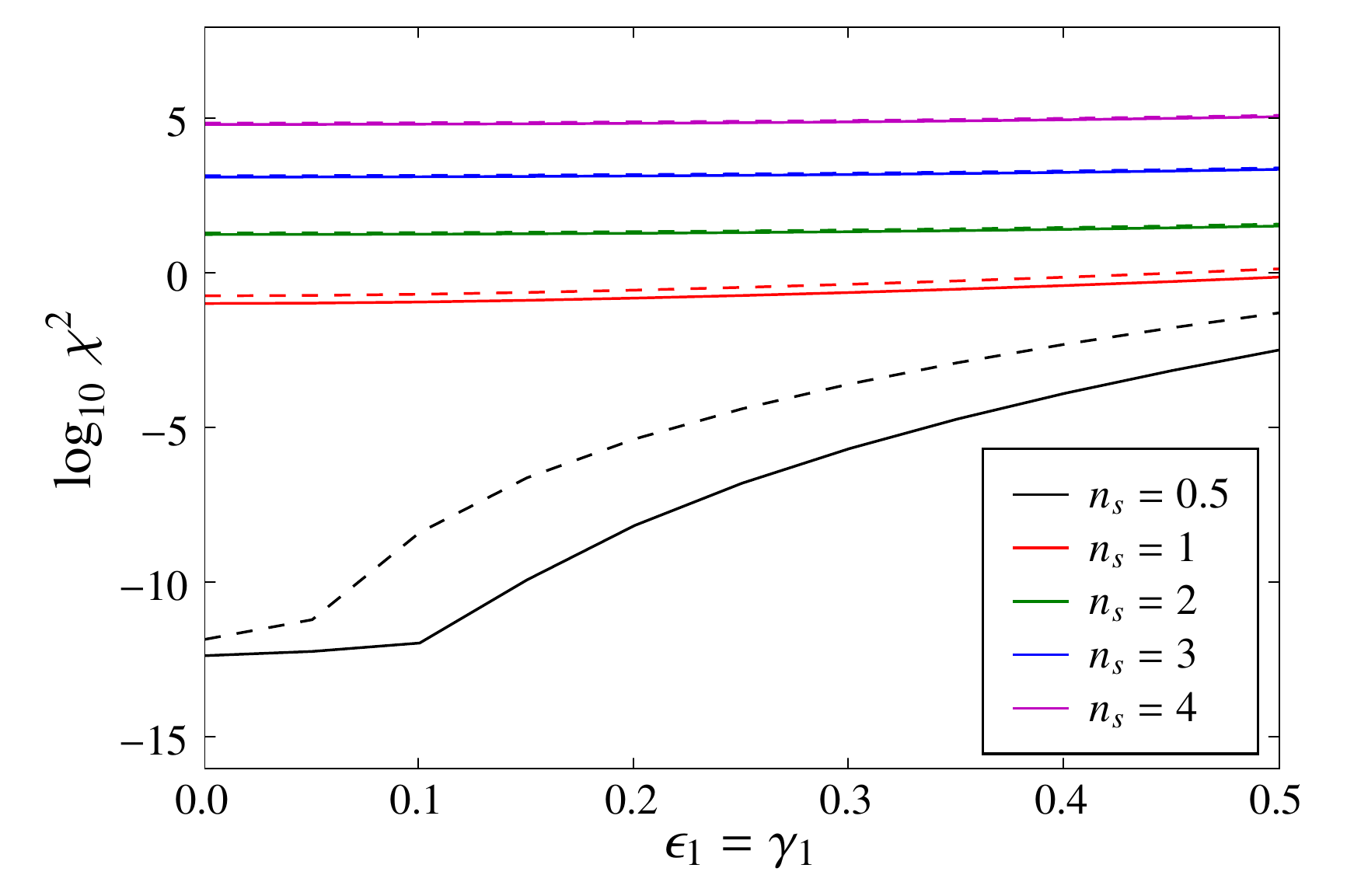}
 \caption{Decomposition goodness-of-fit $\chi^2$ as function of the applied shear for S\'{e}rsic-type galaxy images with $0.5 \leq n_s\leq 4$. The shapelet models use $n_{max}=8$ (dashed lines) or 12 (solid lines). As the images are noise-free, the units of $\chi^2$ are arbitrary.}
 \label{fig:chi2}
\end{figure}

The same trend can be seen quantitatively in Fig. \ref{fig:chi2}, where we show the logarithm of the goodness-of-fit $\chi^2$ of the shapelet models. From this we infer that the modeling errors become stronger with increasing $n_s$ and are almost independent of $\vec\gamma$ for $n_s\geq1$. In other words, while the shapelet decomposition is occupied with modeling a steep profile, it misses most of the ellipticity information. It is worth noting that an increase of $n_{max}$ from 8 to 12 does not lead to substantially lower $\chi^2$, although the number of available modes is raised from 45 to 91. This behavior can be explained by the shape of the higher-order shapelet functions: As they preferentially fit features in the outer regions, they improve the model only in the low-flux regions.

\begin{figure}[t]
 \includegraphics[width=\linewidth]{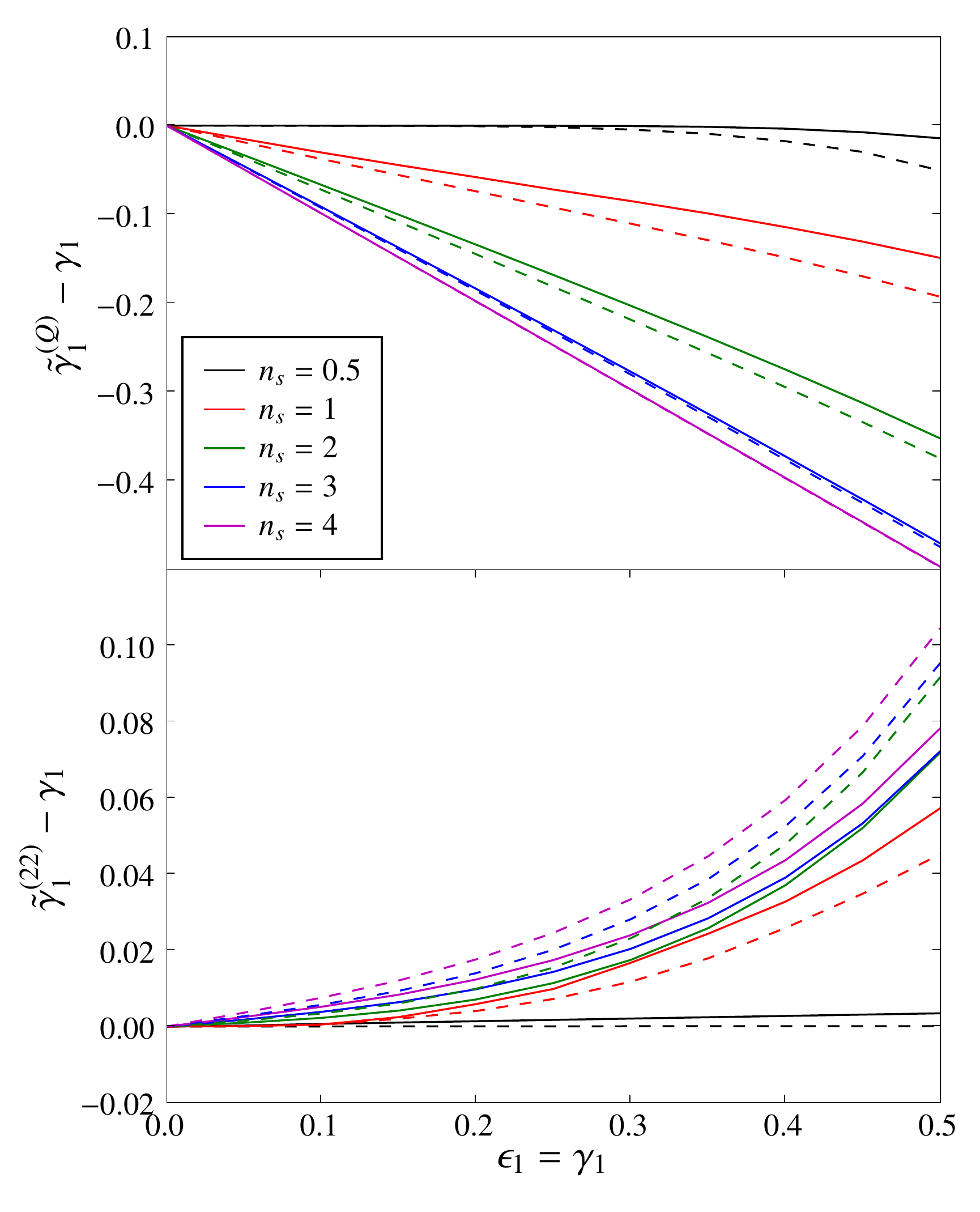}
 \caption{Shear estimates $\tilde{\vec\gamma}^{(Q)}$ (top panel), $\tilde{\vec\gamma}^{(22)}$ (bottom panel) for S\'{e}rsic-type galaxy images as a function of the applied shear. The circular shapelet models use $n_{max}=8$ (dashed lines) or 12 (solid lines).}
 \label{fig:bias-sersic-shear}
\end{figure}

In the top panel of Fig. \ref{fig:bias-sersic-shear}, we show the shear estimator $\tilde{\vec\gamma}^{(Q)}$ as a function of the applied shear. We see that the bias is essentially a linear function of $\gamma_1$ with a negative slope which increases with $n_s$. It is important to note that for $n_s=0.5$ the estimator is unbiased as long as $\gamma$ remains moderate. But, for $n_s\geq 3$ the same estimator is essentially shear-insensitive. Increasing the maximum order $n_{max}$ from 8 to 12 improves the estimator, because the shape at large distances from the center is captured better by the model, and by construction $\tilde{\vec\gamma}^{(Q)}$ makes use of all available orders. But the steeper the profiles becomes the less can high orders contribute to the shear estimation, because the quadrupole moments become dominated by the inner region which is governed by a single central pixel with square shape and therefore vanishing ellipticity\footnote{We will discuss the effects of pixelation in sect. \ref{sec:pixelation}.}. From our prior discussion and Figs. \ref{fig:models} \& \ref{fig:chi2}, we anticipated such a behavior and it clearly confirms our theoretical expectations regarding steep galactic profiles.

In the bottom panel of Fig. \ref{fig:bias-sersic-shear}, the response of the shear estimator $\tilde{\vec\gamma}^{(22)}$ on the same set of galaxies is shown. We can see that the overall bias is mitigated by roughly a factor 4, rather independent of $n_{max}$. This has to be expected as the shapelet basis is orthogonal and thus higher-orders do not change the value of $g^{\prime p}_{2,2}$, which carries the shear signal of $\tilde{\vec\gamma}^{(22)}$. The differences which occur when changing $n_{max}$ are related to a different preferred scale size $\beta$ in the optimization.

As before, galaxies with $n_s=0.5$ can be measured with high fidelity. For steeper profiles the estimator has a -- somewhat surprising -- positive bias, while the shapelet model itself underestimates the ellipticity. We do not fully understand why this estimator overestimates the applied shear, but we can identify two possible reasons:
\begin{itemize}
 \item The estimator has been derived from the action of a infinitesimal shear on a brightness distribution which is perfectly described by a shapelet model \citep{Massey07.2}. In the tests performed here, we intentionally violate these unrealistic assumptions.
 \item Looking at the definition in \veqref{eq:fn2} and the shapelet models in Fig. \ref{fig:models}, we see ring-shaped artefacts for steep profiles, corresponding to radial shapelet modes. Exactly these modes are required for normalizing the estimator. As we know from Fig. \ref{fig:chi2}, the goodness-of-fit -- and hence the abundance of artefacts -- is highly correlated with $n_s$. Thus, the denominator of \veqref{eq:fn2} is probably also plagued by the poor reconstruction quality of steep profiles.
\end{itemize}
We can confirm the last argument by looking at the results of the next higher-order estimator $\tilde{\vec\gamma}^{(42)}$, and find it to be strongly biased and highly unstable under variation of $n_s$. This trend continues for even higher polar order $n$ and renders the family of estimators described by \veqref{eq:fn2} unpredictable and thus unusable for $n>2$. Nevertheless,  $\tilde{\vec\gamma}^{(22)}$ is significantly less biased than $\tilde{\vec\gamma}^{(Q)}$.

\section{Elliptical shapelets}
\label{sec:elliptical_shapelets}

In order to assess the performance of all versions of shapelet image
analysis, we also considered an elliptical implementation of the
Gauss-Laguerre polynomials.  To this purpose, we used a novel code
that we have developed recently (Lombardi et al., in prep.).  This
implementation has many different operating modes regarding a number
of key parameters, such as the truncation mode of the shapelet
decomposition, the fitting of the object shape, and the masking method
used.  However, for sake of simplicity, here we mostly discuss the
results obtained by running the pipeline in a mode that follows
closely the prescriptions of \citet{Bernstein02.1} and \citet{Nakajima07.1}.
We refer to these two papers for all details that are not explicitly mentioned here.

Elliptical shapelets are defined essentially like polar shapelets in a
suitably sheared reference system.  For example, in the simple
case where the elliptical base is oriented along the Cartesian axes,
one can define a new coordinate system $(x'_1, x'_2)$ as
\begin{align}
  x'_1 & {} = x_1 / a \; , & x'_2 & {} = x_2 / b \; ,
\end{align}
where $a$ and $b$ are two scales (the two semi-axes of the base
system). This gives rise to a radial coordinate $r'$, for which we evaluate the shapelet function $B_{n_r,n_l}$ of \veqref{eq:polar_shapelets} on this new system (by keeping $\beta = 1$, since $\beta$
is already encoded in the two scales $a$ and $b$). Generalization for arbitrarily oriented galaxies is straightforward.

All further steps in the decomposition are essentially identical to
the circular shapelet one: for example, the shapelet coefficients are
still evaluated using a $\chi^2$ technique identical to \veqref{eq:chi2}.

The elliptical shapelet decomposition just described, and the
consequent shape estimation, is done using a sequence of individual
steps in our pipeline. 
\begin{enumerate}
\item First, we run SExtractor \citep{Bertin96.1} on the input image (and
  associated weight map, if available) using a rather standard
  configuration.  For each object, among other things, we measure its
  centroid, its shape, and its position angle.
\item The resulting SExtractor parameters of each objects are used as
  initial guesses for the shapelet decomposition.  We first extract
  the images of each object, and mask them using an elliptical mask
  that follows the original SExtractor shape estimate (in particular,
  the mask ellipse has axes that are four times as big as the
  SExtractor ones).  We then use the same SExtractor shape estimate as
  an initial guess for the shapelet elliptical basis.
\item We perform a forward fitting of all pixels contained within the
  mask, and obtain this way a set of shapelet parameters $g^{\prime p}_{n,m}$ and
  associated covariance matrix.  To be able to compare with the circular shapelets, we
  used a \emph{triangular} truncation scheme by requiring that $n = n_r + n_l\le
  n_{max}$, where $n_{max} = 8$ or 12 is the maximum order of the decomposition.
\item The next step is the estimation of the ellipticity.  For this
  purpose, we can either use the quadrupole moments as in \veqref{eq:ellipticity}
  or rely on an iterative approach that we call ``focusing'': we apply
  a sequence of translations, scaling operations, and shear operations
  to the object until it appears perfectly centered, with maximum
  signal to noise, and round.  As shown by
  \citet{Bernstein02.1}, these conditions can be fulfilled by
  requiring that $g^{\prime p}_{1,1} = g^{\prime p}_{2,0} = g^{\prime p}_{2,2} = 0$. 
  % As a further complication, we note that 
  Since we are using an elliptical base for
  our shapelets, the condition $g^{\prime p}_{2,2} = 0$ simply indicates that the
  object is \emph{round} in the elliptical base, i.e. has an ellipticity
  and a position angle that are identical to the ones of the base.
  %Hence, we need to convert this ellipticity estimation into a more
  %natural ellipticity defined in a Cartesian system.
\item Finally, we compare the focusing parameters with the original
  SExtractor one, and we repeat the whole process with the newly
  determined parameters in cases both parameter sets differ significantly.
  This step is typically executed for a relatively small fraction of galaxies only.
\end{enumerate}
%The final shapelet decomposition parameters, with the focusing associated
%ellipticity, are finally saved into a FITS file for further
%processing.

% In case a PSF-corrected decomposition needs to be done, one will
% typically perform a PSF convolution with the shapelet model, and this PSF-convolved model will be then directly used in the $\chi^2$ minimization.

A key point of the whole process is the presumed independence of the
final result on the input SExtractor parameters.  Of course, since the
ellipticity estimation involves two non-linear iterative approaches
(one for the design of the shapelet base and image mask, and a second
for focusing), we cannot expect a perfect independence.  In other
words, if we start from completely wrong source centroids and shapes,
it will be very unlikely that we will be able to perform decent
decompositions and that the whole pipeline will converge
towards sensible results.  On the other hand, if the method relies too
much on ``good'' initial parameters, and does not tolerate relatively
small errors on the SExtractor catalog, we will likely face severe
problems, because although SExtractor has a very good performance in
common cases, it clearly has not been designed with weak lensing
studies in mind, and does not reach the accuracy needed in this field.

\subsection{Results}
In order to test the ability of our elliptical shapelet pipeline to
deal with inaccurate input catalogs, we biased the input ellipticity
by $20\%$ toward circular objects, and measured the residual bias
left in the recovered ellipticity $\tilde{\gamma}^{(el)}$. 

\begin{figure}[t]
 \includegraphics[width=\linewidth]{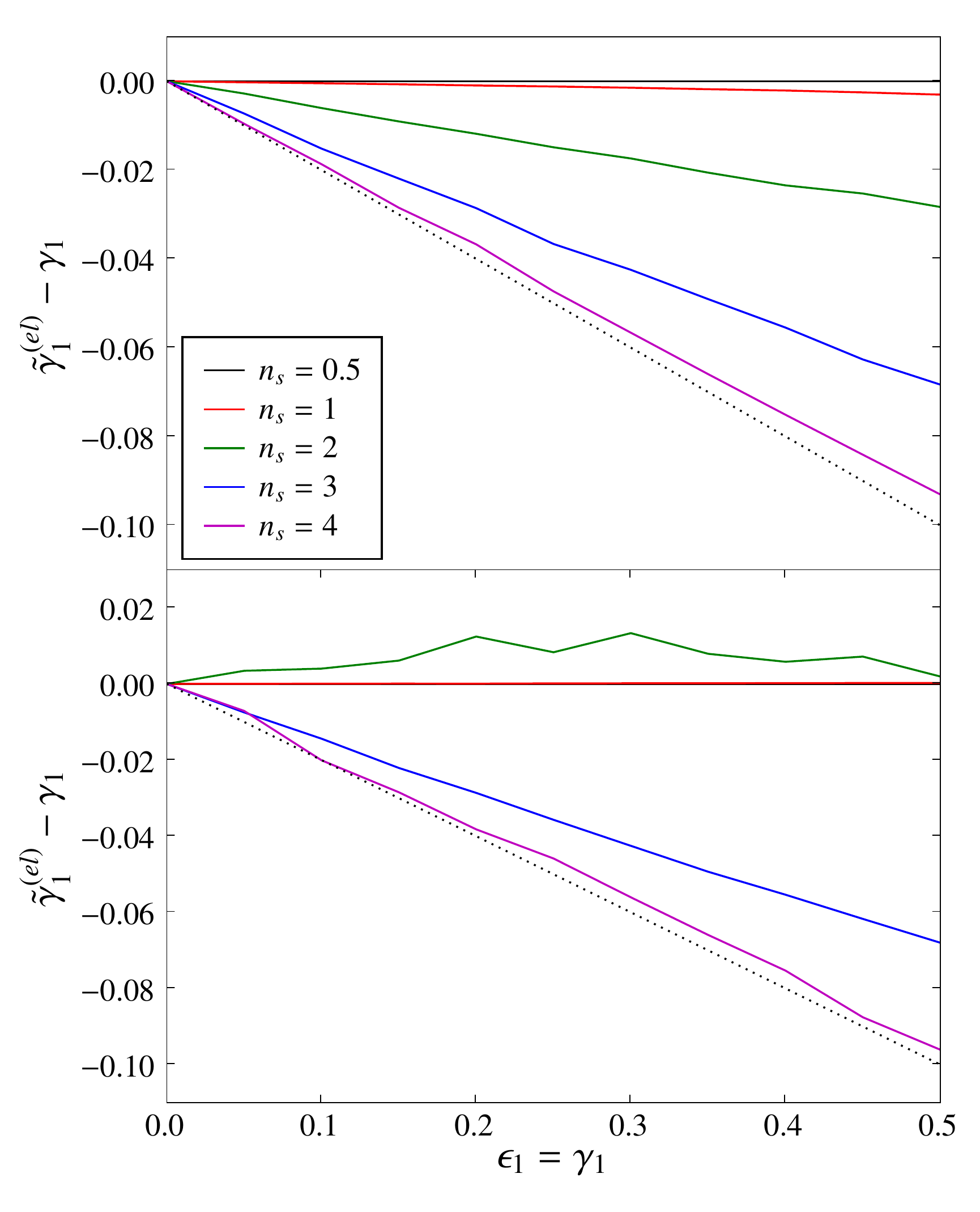}
 \caption{Shear estimate $\tilde{\vec\gamma}^{(el)}$ for S\'{e}rsic-type galaxy images as a function of the applied shear. The elliptical shapelet models use $n_{max}=8$ (top panel) or 12 (bottom panel). The dotted line in both panel represents the bias of -20\%, which we artificially applied to the initial guess of the ellipticity.}
 \label{fig:elliptical-nmax}
\end{figure}

As shown by Fig. \ref{fig:elliptical-nmax}, this test
produced conceptually similar results as with circular shapelets: 
for small S\'{e}rsic indices, we could recover the true ellipticity without any 
significant bias, while the estimates degrade significantly as we approach $n_s = 4$.  
The situation is improved when $n_{max}$ is raised, because the transformations done during the focusing step take all available orders into account, in contrast to the simpler description underlying the construction of $\tilde\gamma^{(n2)}$ in \veqref{eq:fn2}. With $n_{max}=12$ (bottom panel of Fig. \ref{fig:elliptical-nmax}), galaxies with $n_s=0.5$ and 1 have shear estimates without bias. When raising $n_s$ beyond that, the bias is at first positive before it becomes negative. Investigating this feature more closely, we find this estimator to have an oscillatory tendency once the bias sets in.

In summary, it appears that the pipeline is unable to fully correct inaccuracies
in the SExtractor catalog for galaxies showing a profile close to the
De Vaucouleurs one. 

% This behavior can be understood by recalling that the shapelet
% decomposition is performed using Laguerre polynomials truncated by a
% Gaussian one.  As a result, if the galaxy profile differs
% significantly from the Gaussian profile, shapelets are very
% inefficient, and a good representation of the object is not reached
% until we go to an exceedingly large order.  Incidentally, a similar
% behavior is expected in the representation of Point Spread Functions
% (PSFs) with large tails (often present in space-born telescopes).

\section{Observational systematics}
\label{sec:systematics}

In more realistic simulations or observational data, the galactic shapes are recorded after convolution with the PSF, pixelation by the CCD, and degraded by pixel noise. We now discuss the impact of these effects on shear estimation with shapelets.

\subsection{PSF convolution}
\label{sec:convolution}

Clearly, a convolution creates shallower profiles which can be better described by shapelet models. Therefore, the typical goodness-of-fit values, in particular for steeper profiles, are considerable lower than in the unconvolved case. If the PSF shape is perfectly described by its shapelet model, one can exactly undo a convolution in shapelet space. In such a case, the shape obtained by deconvolving a PSF-convolved galaxy model must approximate the true, unconvolved shape $G^\prime$ better than its direct model $\tilde{G}^\prime$\footnote{For our argument, we only exploit that convolution with the PSF renders observed profiles shallower. Hence, the last statement is probably still true for imperfect shapelet models of the PSF -- which are likely to introduce other systematics.}.

To verify this new hypothesis, we convolved S\'{e}rsic-type galaxies in pixel space with PSF shapes $P$ obtained from shapelet models\footnote{In the terminology of this work, that means $P=\tilde{P}$.},
\begin{equation}
 C = P \star G^\prime.
\end{equation}
For circular shapelets, $C$ is modeled with shapelets and explicitly deconvolved from $\tilde{P}$ in shapelet space, while for the elliptical shapelets we obtain the unconvolved shape by convolving the model with the PSF and fitting the outcome to the image data.

\begin{figure}[t]
 \includegraphics[width=\linewidth]{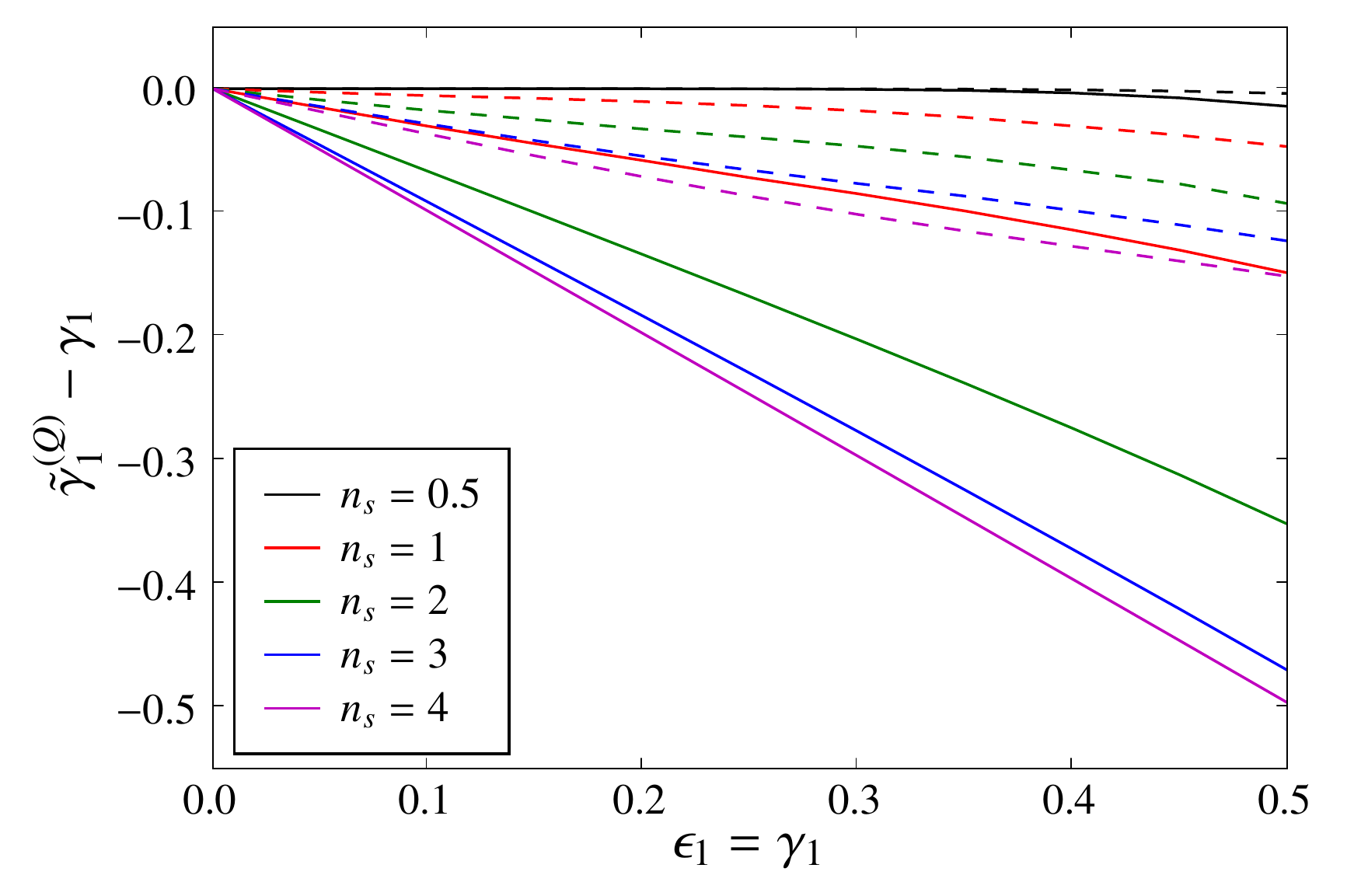}
 \caption{Similar to Fig. \ref{fig:bias-sersic-shear}: Solid lines show bias on shear estimate $\gamma^{(Q)}$ from circular shapelet models with $n_{max}=12$ for unconvolved galaxy images, while dashed lines are obtained for the same set of galaxies after convolution with a Gaussian of 5 pixels FWHM.}
 \label{fig:bias-convolution}
\end{figure}

In Fig. \ref{fig:bias-convolution} we compare the bias of the shear estimates from unconvolved galaxies images and from the same set of galaxies after convolution with a Gaussian PSF with a FWHM of 5 pixels -- thus FWHM $= R_e$. As expected, the overall shape of the convolved images -- and thus also their unconvolved shapes -- can be better modelled with shapelets. Hence, the shear estimator $\tilde\gamma^{(Q)}$ profits manifestly because it makes use of the entire shape information: its bias is lowered by a factor 3. We observed similar improvements also for $\tilde\gamma^{(22)}$ and $\tilde\gamma^{(el)}$. 

We varied the size of the PSF and found that in general the bias is lower for larger PSFs. Additionally, we found $\tilde\gamma^{(22)}$ and $\tilde\gamma^{(el)}$ to be rather sensitive to changes of $n_s$ and FWHM$/R_e$. Finally, the bias of all estimators also depends in a non-trivial way on the shape of the PSF, because the fidelity of the model for the convolved galaxy image determines the accuracy of the estimators.

\subsection{Pixelation}
\label{sec:pixelation}

Images from CCDs are obtained by collecting the light within pixels of approximately square shape.
For measuring shapes, pixelation has important consequences. If the size of the object is small compared to the pixel size, we cannot describe the true continuous shape of the object but rather its piecewise approximation with pixel-sized step functions. Modeling approaches like the shapelets method can take pixelation into account by integrating the model values within the pixels \citep{Massey05.1}. In case of convolved images, the deconvolution procedures also treat pixelation consistently, if the PSF shape has been measured from images with the same pixelation \citep[e.g.][]{Bridle08.1}.\\
For estimating the shear, an additional problem is of relevance. As the smallest piece of information within an image is given by a single pixel with square shape, we can only infer shear information from an object for which we can measure more than a single pixel. Particularly for galaxies with steep profiles, the largest fraction of the flux is registered in the pixel which is closest to the centroid. Then, also the shear information is dominated by this central pixel, which does not have any preferred direction, and hence biased low.

% \begin{figure}[t]
%  \includegraphics[width=\linewidth]{bias-pixelation}
%  \caption{Multiplicative bias $m$ as a function of pixel size for $\gamma^{(Q)}$ (top), $\gamma^{(22)}$ (middle), and $\gamma^{(el)}_{-20\%}$ (bottom). An effective smaller pixel size is achieved by raising the half-light radius $R_e$ of $G$. The dotted line in the bottom panel represents the bias on the ellipticity prior.}
%  \label{fig:bias-pixelation}
% \end{figure}

For this work, it is important to verify that the biases related to steep profiles are not entirely a pixelation problem, but stem from the shape mismatch as discussed before. Therefore, we also made sets of images with $R_e = 10,\,20$.\\
Although there are some differences between the three tested estimators, we found a common trend when increasing the size of the galaxies: The results for galaxies with $n_s\leq 1$ are essentially unchanged. In particular, the bias does not vanish when the side length of a pixel is reduced to one quarter. This shows that there is a profile dependent bias, independent of pixelation.
The estimates for steeper profiles benefit from smaller pixel sizes, but the results still remain more strongly biased than for shallower profiles.

\subsection{Pixel noise}
\label{sec:pixel_noise}

There is an additional effect related to the discussion of pixelation: In the presence of pixel noise, a smaller number of significant pixels remains for each galaxy. In particular, steep profiles therefore tend to be reduced to some pixels or even only a single pixel close to center of the galaxy, which is then fitted by the model. Thus, we expect pixel noise to behave in a similar way as strong pixelation.

We performed the same tests again with realistic noise added to $G^\prime$, but apart from additional statistical uncertainty it did not lead to different results. In fact, we can confirm that steep profiles are affected more strongly by pixel noise than shallower ones.

There is another important point to note: Increased pixel noise would normally lead to a lower $n_{max}$, as the shapelet models are typically tuned such that they do not or only marginally pick up noise fluctuations. Reducing the maximum order typically leads to more prominent modeling problems, as we already discussed before and consequently to poorer results from most shear estimators.

\section{Conclusions}
\label{sec:conclusions}

\begin{itemize}
 \item Shear estimates from circular shapelets are biased if the shape to be described has too steep a profile (steeper than a Gaussian) or too large an ellipticity. Profile mismatch is the more important source of bias.
\item For elliptical shapelets, profile mismatch still poses a considerable problem, because the shapelet models cannot fully correct biases of the ellipticity prior when the profile becomes steeper than exponential.
\item Different shear estimators can mitigate the bias, but never eliminate it completely because the shape mismatch generally affects all shapelet modes.
\item For flexion estimates we are even more concerned. While the overall octupole power might be lower than the quadrupole power, the bias introduced by the shape mismatch becomes more important for higher-order moments.
\item Convolution with a PSF renders all observable shapes shallower and allows the treatment of pixelation, and hence facilitates a more accurate description by shapelet models.
\item The ellipticity estimates could be calibrated provided we have simulations detailed enough to properly describe the distribution of intrinsic galaxy morphologies and the effect of PSF convolution on it. Furthermore, we would need to be able to infer the appropriate calibration parameters from observations. To achieve the latter is at least as hard as measuring the ellipticity.
\item Even if an effective calibration for an entire galaxy ensemble could be achieved, it does not ensure unbiased results when the ensemble is sliced (e.g. in redshift bins for weak-lensing tomography or spatially for cluster-mass reconstructions).
\item Lensing simulations with shapelet source models \citep[e.g.][]{Massey04.1,Meneghetti08.1} systematically favor shapelet pipelines for weak-lensing measurements as the shapelet basis forms the optimal way of shape description. 
\item The mentioned shortcomings are mostly relevant for small and faint galaxies which are described by shapelet models with low $n_{max}$. For objects with a high signal-to-noise ratio, modeling with high $n_{max}$ reduces these effects. Additionally, the artefacts which can occur on bright galaxies have a similar impact on  shapelet models for galaxies which are intrinsically similar. We conclude therefore that morphological investigations of galaxy populations are still feasible with shapelets. However, measures of e.g. their ellipticity or profile slope distribution are likely biased.
\item PSF shapes typically follow the Moffat profile \citep{Moffat69.1} $p_m\propto(1+\alpha r^2)^{-\beta_m}$ and can thus be very well modeled already with a pure Gaussian -- which is the limiting case for $\beta_m\rightarrow\infty$ \citep{Trujillo01.1}. In particular for PSFs from ground-based observations with its at most mild ellipticity, shapelets can reach excellent modeling accuracy.
\end{itemize}

\section*{Acknowledgments}
PM is supported by the DFG Priority Programme 1177. This work has gained substantially from the ideas, suggestions and criticism of Massimo Meneghetti, Ren\'e Andrae, Konrad Kuijken and Ludovic van Waerbeke.

\end{document}